# A single low-energy, iron-poor supernova as the source of metals in the star SMSS J031300.36-670839.3

S. C. Keller$^{\backslash}$, M. S. Bessell$^{\backslash}$, A. Frebel$^{*}$, A. R. Casey$^{\backslash}$, M. Asplund$^{\backslash}$, H. R. Jacobson$^{*}$, K. Lind$^{*}$, J. E. Norris$^{\backslash}$, D. Yong$^{\backslash}$, A. Heger$^{+}$, Z. Magic$^{\triangle\backslash}$, G. S. Da Costa$^{\backslash}$, B. P. Schmidt$^{\backslash}$, & P. Tisserand$^{\backslash}$

**The element abundance ratios of four low-mass stars with extremely low metallicities indicate that the gas out of which the stars formed was enriched in each case by at most a few, and potentially only one low-energy, supernova[1,2,3,4]. Such supernovae yield large quantities of light elements such as carbon but very little iron. The dominance of low-energy supernovae is surprising, because it has been expected that the first stars were extremely massive, and that they disintegrated in pair-instability explosions that would rapidly enrich galaxies in iron[5]. What has remained unclear is the yield of iron from the first supernovae, because hitherto no star is unambiguously interpreted as encapsulating the yield of a single supernova. Here we report the optical spectrum of SMSS J031300.36-670839.3, which shows no evidence of iron (with an upper limit of $10^{-7.1}$ times solar abundance). Based on a comparison of its abundance pattern with those of models, we conclude that the star was seeded with material from a single supernova with an original mass of ~60 M$_\odot$ (and that the supernova left behind a black hole). Taken together with the previously mentioned low-metallicity stars, we conclude that low-energy supernovae were**

$^{\backslash}$ Research School of Astronomy and Astrophysics, Australian National University, ACT 2601, Australia
$^{*}$ Dept. of Physics, Massachusetts Institute of Technology & Kavli Institute for Astrophysics and Space Research, Cambridge, MA 02139, USA
$^{*}$ Institute of Astronomy, University of Cambridge, Madingley Road, Cambridge CB3 0HA, United Kingdom
$^{+}$ Monash Centre for Astrophysics, School of Mathematical Sciences, Monash University, Victoria 3800, Australia.
$^{\triangle}$ Max-Planck-Institut für Astrophysik, Karl-Schwarzschild-Str. 1, 85741 Garching, Germany

**common in the early Universe, and that such supernovae yield light element enrichment with insignificant iron. Reduced stellar feedback both chemically and mechanically from low-energy supernovae would have enabled first-generation stars to form over an extended period. We speculate that such stars may perhaps have had an important role in the epoch of cosmic reionization and the chemical evolution of early galaxies.**

Whereas the solar spectrum contains many thousands of spectral lines due to iron and other elements, the high-resolution (R = 28,000) optical spectrum SMSS J031300.36-670839.3 (hereafter SMSS 0313-6708) is remarkable for the complete absence of detectable iron lines. Figure 1 shows a portion of the spectrum that possesses signal-to-noise = 100 per resolution element in the vicinity of one of the strongest iron lines (Fe I at 385.9 nm). The non-detection of iron lines places an upper limit on the iron abundance of the star, [Fe/H] < -7.1 ([A/B] = $\log_{10}(N_A/N_B) - \log_{10}(N_A/N_B)_\odot$ where $N_A/N_B$ is the number ratio of atoms of elements A and B and the subscript $\odot$ refers to the solar value) at a 3σ confidence level. This upper limit is 30 times lower than the iron abundance in HE 1327-2326 with [Fe/H] = -5.6[2], the most iron-deficient star previously known.

The paucity of absorption lines in the spectrum of SMSS 0313-6708 allows us to derive the abundance of only four chemical elements. The calcium abundance is determined to be [Ca/H] = -7.0. Given that existing studies have shown that [Ca/Fe] = +0.4 for the majority of extremely metal-poor stars[6], the [Ca/H] value would be consistent with an extraordinary low iron abundance limit. We suggest below, however, that the Ca abundance of the star has its origin in a specific process and no coupling to the Fe abundance is expected. Magnesium ([Mg/H] = -3.8) and carbon ([C/H] = -2.6), however, are both highly enhanced relative to iron and calcium (see Figure 2). Table 1 presents the details of our chemical abundance analysis procedure and derived abundances and abundance limits. Figure 3 presents the abundance pattern of SMSS 0313-6708.

For the most iron-poor stars the degree of chemical diversity increases with decreasing metallicity[6], indicative of the diversity of the few supernovae that enriched the gas prior to their formation. Yields from supernovae with explosion energies typical of those seen in the nearby universe are required to explain the observed abundances in stars with [Fe/H] > -3.5 except for a few elements[6]. For the most iron-poor stars, however, the supernovae required to explain the chemical abundances suggest unusual low explosion energy black hole-forming events[7].

Stochastic galaxy chemical evolution models that invoke low-explosion energy supernovae[8] with a burst of Population III star formation over 25 > z > 17, reveal that stars such as SMSS 0313-6708 are the result of star formation in gas enriched by only one Population III supernova. Whereas more iron-rich stars ([Fe/H] > -4.5) may be interpreted as composites of supernovae yields, galaxy chemical evolution models show predict stars with the iron abundance of SMSS 0313-6708 follow from a single supernova event.

We have compared the abundance pattern of SMSS 0313-6708 to the nucleosynthetic yields of model Population III supernovae that range in progenitor mass, explosion energy and internal mixing[9]. We find that a $1.8 \times 10^{51}$ erg explosion of a 60 $M_\odot$ star of primordial initial composition with a small amount of ejecta mixing due to Rayleigh-Taylor instabilities[10] is the optimal match to the observed abundance pattern (see Figure 3). In this model, a central black hole is formed into which the core of the massive star is subsumed. The extensive fallback of material into the black hole traps the centrally-located iron and other heavy elements synthesised during the star's lifetime. Lighter elements, for example carbon and magnesium, residing at larger radii within the supernova progenitor are dispersed in the explosion.

The observed abundance pattern of SMSS 0313-6708 does not support supernova progenitors outside the range of 10 – 70 $M_\odot$. Supernovae less than 10 $M_\odot$ release large amounts of iron, those greater than 70 $M_\odot$ do not produce the observed carbon enhancement and lead to excessive nitrogen[9]. In particular, pair-instability supernovae are expected to yield [C/Fe] ~ 0 compared to the [C/Fe] > 4.5 observed[10].

Our observations argue against a significant non-axisymmetric supernova, such as one in which much of the explosion energy is channelled into a jet, since this would entrain and eject material from the core region. Any appreciable jet would lower [Mg/Fe] below that observed[11]. Furthermore, the relatively low nitrogen abundance, implied by our upper limit compared to carbon, suggests a slowly rotating progenitor star[12].

Interestingly, the calcium released in our model of a 60 $M_\odot$ progenitor is not synthesized in the supernova explosion itself; rather it is produced during the stable hydrogen-burning phase. In the metal-free progenitor, thermal equilibrium is obtained only once the core of the star obtains significantly higher temperature and density compared with metal-rich stars. Under such conditions the triple-alpha process enables the synthesis of small amounts of carbon, nitrogen, and oxygen that subsequently catalyse the CNO energy production cycle[13]. Calcium production is the result of breakout from the CNO-cycle[14]. As a consequence we expect the calcium abundance to be decoupled from those of heavier elements in the most metal-poor stars. Such synthesis is only appreciable in initially metal-free stars where internal temperatures are sufficiently elevated.

With SMSS 0313-6708, there are now five stars known with [Fe/H] < -4.5. The metallicity distribution function of stars reflects the integrated iron production from previous generations of supernovae. A simple model of instantaneous gas mixing[15] shows that the number of stars at a given low metallicity drops by a factor of ten for each factor of ten reduction in [Fe/H]. This simple model proves appropriate[6] for [Fe/H] > ~-4. With the addition of SMSS 0313-6708, the five most iron-poor stars span a range in iron abundance of at least 2.5 decades in metallicity. The probability of such an observed metallicity density function arising from the Simple Model by chance is low (< 1.6%; assuming they are drawn from a completeness-corrected sample normalised to the known 30 stars with -4.0 < [Fe/H] < -3.5). This is suggestive that for [Fe/H] < -4.5 the assumption of instantaneous mixing of star-forming gas no longer holds.

The rate of mixing and enrichment from Population III stars has an important impact on the epoch of reionization. Studies have highlighted that z = 6 – 8 galaxies do not produce sufficient ionizing photons to lead to the reionization of the intergalactic medium[16]. With a mass function dominated by massive stars, Population III stars produce ~10x the ionizing radiation over their lifetimes compared with Population II stars that constitute the observable z = 6 – 8 galaxies. Prodigious radiation output by Population III stars at z ~ 20 offers a potential resolution to the ionizing radiation shortfall.

However, simulations of the early universe that include energetic pair-instability supernovae result in very rapid mixing of supernova ejecta leading to the pollution of large volumes of gas with metals. This rapidly terminates Population III star formation, and hence Population III is extinguished before it contributes significant ionizing radiation[17].

On the other hand, supernovae with low explosion energy and low metal yield, which our observations reveal, result in slow mixing and enrichment of the interstellar medium. This enables the formation of Population III stars over a protracted time, thus allowing the population to release substantial ionizing radiation to potentially account for the apparent radiation shortfall, although we are unable to quantify this at present.

Methods summary

SMSS 0313-6708 is located at RA = 03:13:00.4, Dec = -67:08:39 (equinox 2000) and has an apparent visual magnitude $V$ = 14.7. It was discovered in the on-going SkyMapper Southern Sky Survey[18]. Stellar parameters were determined from low-resolution spectrophotometry to be effective temperature $T_{eff}$ = 5,125 K, and surface gravity log g = 2.3 [cgs] (Extended Data Figure 1). We adopt a microturbulent velocity of 2.0 km/s (the derived abundances are not sensitive to this choice). The temperature and gravity are consistent with the stellar hydrogen line profiles and the derived lithium abundance (Extended Data Figure 2). The uncertainty in temperature is of order 100K, and for surface gravity it is 0.2. Two model atmospheres are considered as the basis for abundance analysis; a Castelli – Kurucz 1-d hydrostatic model[19], and a spatially and temporally averaged 3-d hydrodynamical model from the Stagger-grid[20] which we denote <3-d>. Corrections for departures from local thermodynamic equilibrium (NLTE) are computed for the <3-d> model following [21]. The abundances of Li, C and upper limits for N, and O have been derived from spectrum synthesis of the lithium 670.7 nm doublet, carbon G-band, NH band

(336.0 nm), and [OI] at 630 nm, respectively. Apart from the molecular features due to CH, only one Ca II (393 nm), one Li I,(671 nm) and 5 Mg I (382.9, 383.2, 383.8, 517.2, and 518.4 nm) lines are detectable in our high S/N spectrum. Solar abundances for elements are from [22]. The upper limit to the iron abundance in the star (Figure 1 & Extended Data Figure 3) was determined through Markov Chain Monte Carlo analysis of addition of the strongest iron lines.

**Acknowledgements** Australian access to the Magellan Telescopes was supported through the National Collaborative Research Infrastructure Strategy of the Australian Federal Government. A.H., M.A., M.S.B., A.R.C., G.D.C., S.K., J.E.N., and D.Y. acknowledge the support of Australian Research Council (grants FT120100363, DP120101237, DP0984924, DP0878137, and LF0992131). A.F. acknowledges support from NSF grant AST-1255160. A.R.C acknowledges the support from the Australian Prime Minister's Endeavour Award Research Fellowship. K.L. acknowledges support from the European Union FP7 programme through ERC grant number 320360.

**Author Contributions** SkyMapper telescope was developed by B.P.S., G.D.C., M.S.B., P.T., and S.K. The SkyMapper data reduction procedure required to provide calibrated photometry from which the star was drawn was developed by S.K.. M.S.B. obtained the intermediate-resolution spectrum and drew the target to the team's attention. H.R.J., A.R.C., A.F. and S.K. obtained the high-resolution spectrum of the target, reduced the data and performed the chemical abundance analysis utilizing the spectral analysis package developed by A.R.C. The MCMC calculations to provide the upper limit to [Fe/H] were performed by A.R.C. K.L. performed NLTE calculations, Z.M. and M.A. constructed the <3-d> atmosphere models, and A.H. the SNe models. B.P.S., A.H. and D.Y. contributed to SNe yields and MDF analysis. All authors discussed the results and commented on the manuscript.

**Author information** Reprints and permissions information is available at www.nature.com/reprints. The authors declare that they have no competing financial interests.

**Table 1  Chemical abundances of SMSS 0313-6708**

| Element | $[X/H]_{1-d, LTE}$ | $[X/H]_{<3-d>}$ |
|---------|--------------------|-----------------|
| Li I    | 0.7$^{\backslash}$ | 0.7$^{\backslash}$ |
| C (CH)  | -2.4               | -2.6$^a$        |
| N (NH)  | <-3.5              | <-3.9$^a$       |
| O I     | <-2.3              | <-2.4$^a$       |
| Na I    | <-5.5              | <-5.5$^b$       |
| Mg I    | -4.3               | -3.8$^b$        |
| Al I    | <-6.2              |                 |
| Si I    | <-4.3              |                 |
| Ca II   | -7.2               | -7.0$^b$        |
| Sc II   | <-5.0              |                 |
| Ti II   | <-6.3              |                 |
| V II    | <-3.3              |                 |
| Cr I    | <-6.3              |                 |
| Mn I    | <-5.8              |                 |
| Fe I    | <-7.3              | <-7.1$^b$       |
| Co I    | <-4.9              |                 |
| Ni I    | <-6.4              |                 |
| Cu I    | <-3.5              |                 |
| Zn I    | <-3.4              |                 |
| Sr II   | <-6.7              |                 |
| Ba II   | <-6.1              |                 |
| Eu II   | <-2.9              |                 |

Table 1 Details the abundance ratios for SMSS 0313-6708 as derived from our Magellan/MIKE spectra. Typical (1σ) observational uncertainties in the quoted abundances are 0.1 decades in metallicity, except in the case of C and N where 0.2 is appropriate. $^{\backslash}$Lithium abundance is expressed as A(Li) = $\log_{10}$(N(Li))/N(H) + 12. The abundances are based either on <3-d>, LTE (marked with $^a$) or <3-d>, NLTE (marked with $^b$) calculations respectively.

**Figure 1. A comparison of the spectrum of SMSS 0313-6708 to that of other extremely metal-poor stars.** Metal-poor stars of similar temperature and surface gravity are chosen from the literature. The spectrum of SMSS 0313-6708 shows an absence of detectable Fe I lines (panel a) and is dominated by molecular features of CH (panel c). Panel (b) shows the vicinity of what should be one of the strongest iron lines in the UV/optical wavelength region. Overlaid are synthesised line profiles (1-d LTE) for [Fe/H]=-7.5, -7.2 (solid) and -6.9.

**Figure 2. A comparison of the abundance ratios observed in SMSS 0313-6708 with those of other extremely metal-poor stars.** Panel (a) shows [Mg/Ca] as a function of [Fe/H]. The dashed line is the Solar abundance ratio. In panel (b) we see [C/H] as a function of [Fe/H] where dashed lines show lines of [C/Fe]. Panel (c) shows [O/H] against [Fe/H]. Our target is marked as the star symbol. Extremely metal-poor carbon-enhanced stars from [23] appear as solid circles. Indicative 1 s.d. error bars are shown. For consistency with the data from Table 4 of [23] the 1-d LTE abundances for SMSS 0313-6708 are plotted.

**Figure 3. The elemental abundance pattern for SMSS 0313-6708.** Determined abundances are shown in solid symbols (observational uncertainties are smaller than symbol size). Due to the paucity of observable absorption lines the majority of abundances are 3σ upper limits (open circles). The solid line shows the abundances predicted for a 60 $M_\odot$ Population III star of relatively low explosion energy ($1.8\times10^{51}$ erg) and low levels of internal mixing (see [10] for details). The dashed line shows the expected yield from a 200 $M_\odot$ supernova (pair-instability

mechanism). Such a massive progenitor leads to a [Mg/Ca] ratio that is much lower than what is observed.

**Extended Data Figure 1. The summary of spectrophotometric analysis of SMSS 0313-6708.** In the top frame the blue line shows the observed spectrum and the green line is the best-fitting model spectrum. The red line beneath shows the residual spectrum. In the lower frame the cross-hair marks the location of the best-fitting Teff and log g and the rms values of the fit are represented in color for a subgrid. Halo isochrones are shown to assist in the selection of the most likely parameters. The interstellar reddening is found to be E(B-V)=0.04.

**Extended Data Figure 2. The comparison of the hydrogen-beta Balmer line profile of SMSS 0313-6708 to literature stars of bracketing stellar parameters.** We utilise this comparison as a qualitative verification of the stellar parameters for SMSS 0313-6708 determined from spectrophotometric analysis. Literature values for $T_{eff}$ and log $g$ respectively are stated in the top panel adjacent to the star identifier.

**Extended Data Figure 3. Determination of upper limits to the iron abundance.** The black line with associated uncertainties is the observed spectrum stacked in the vicinity of the strongest iron lines. The colored lines show 67.8% (blue), 95% (magenta), and 99.7% (red) confidence upper limits for [Fe/H]$_{1D,NLTE}$.

**METHODS**

**Discovery.** SMSS 0313-6708 is located at RA = 03:13:00.4, Dec = -67:08:39 (equinox 2000) and has an apparent visual magnitude $V$ = 14.7. On the basis of photometry obtained with the SkyMapper telescope[18] on October the 2$^{nd}$ 2012, the star was predicted to be of particularly low metallicity. The SkyMapper telescope utilizes a novel filter set that places tight constraints on the fundamental properties of the stars surveyed, namely effective temperature, surface gravity and stellar metallicity.

**Abundance analysis.** Two model atmospheres are considered as the basis for abundance analysis; a Castelli – Kurucz 1-d hydrostatic model[19], and a spatially and temporally averaged 3-d hydrodynamical model from the Stagger-grid[20] which we denote <3-d>. Corrections for departures from local thermodynamic equilibrium (NLTE) are computed for the <3-d> model following [21]. The abundances of Li, C and upper limits for N, and O have been derived from spectrum synthesis of the lithium 670.7 nm doublet, carbon G-band, NH band (336.0 nm), and [OI] at 630 nm, respectively. Apart from the molecular features due to CH, only one Ca II (393 nm), one Li I,(671 nm) and 5 Mg I (382.9, 383.2, 383.8, 517.2, and 518.4 nm) lines are detectable in our high S/N spectrum. Solar abundances for elements are from [22]. Observational uncertainties in the derived abundances (Table 1) are 0.1 decades in metallicity. The local signal-to-noise in the observed spectrum governs the uncertainty. In the case of upper limits to abundance these were determined by matching synthesized spectra to the local average minima in the vicinity of the line in question. Observational uncertainties are larger in the case of C and

N (σ = 0.2 decades in metallicity) where synthetic spectra are compared to the multiple features of molecular bands.

**Medium-resolution spectroscopy.** The star's extremely metal-poor status was confirmed through medium-resolution spectroscopy with the WiFeS spectrograph[24] on the ANU 2.3 m telescope on the 2nd January 2013.

**High-resolution spectroscopy**. We observed SMSS 0313-6708 on five different nights (namely 23-25th January and the 6-7th February 2013) with the Magellan Inamori Kyocera Echelle (MIKE) spectrograph[25] at the 6.5 m Magellan Clay telescope, Chile. All observations were taken using a 0.7" slit, providing a spectral resolution of R = 35,000 in the blue arm and R = 29,000 in the red arm. Calibration frames were taken at the start of each night, including twenty flat-field frames (10 quartz, 10 milky) and 10 Th-Ar arc lamp exposures for wavelength calibration. Data reduction utilized the Carpy data reduction pipeline[26]. Each reduced echelle order was carefully normalized using a third order spline with defined knot spacing. Normalized orders were stitched together to provide a single one-dimensional spectrum from 336-940 nm.

**Adopted stellar parameters.** The WiFeS / ANU 2.3m spectrum was flux calibrated by comparison with spectrophotometric standard stars as described in [27]. Model atmosphere fluxes [28] were then compared to those observed from SMSS 0313-6708 as described in [29] to determine the best match to the spectrum. Extended Data Figure 1 shows the result of the fit to SMSS 0313-6708. We have adopted the $T_{eff}$ and log $g$ resulting from this analysis in the subsequent abundance analysis: Teff = 5,125 K, and surface gravity log g = 2.3 [cgs]. We adopt a microturbulent velocity of 2.0 km/s (the derived abundances are not sensitive to this choice)..

To confirm that the values utilised are appropriate, we have made a comparison between the Balmer line profiles observed and those of stars with bracketing $T_{eff}$ and log $g$ observed with the same instrument and settings (seen in Extended Data Figure 2 a, b). The uncertainty in temperature is of order 100K, and for surface gravity it is 0.2.

Lines of interstellar absorption of Ca-K and Na-D are apparent in the spectrum. We determine the equivalent width of the Na-D1 line to be 10.9 pm. The relation between equivalent width and interstellar reddening from [30] provides E(B-V) = 0.04 ± 0.01 which we adopt. Using the infra-red flux method for the *BVJHK* system[31] (*BV* photometry from APASS[32]; *JHK* from 2MASS[33]) we derive a photometric temperature of $T_{eff}$ = 5,210 ± 64 K. From utilisation of *griJHK*[31] (*gri* from APASS) we determine $T_{eff}$ = 5,265 ± 76 K. These values and those determined through spectrophotometric analysis are in concordance.

**Lithium.** Lithium is detected in the star at a level A(Li) = $\log_{10}$(N(Li))/N(H) + 12 = 0.7 that displays significant depletion from the primordial Big Bang level (A(Li) = 2.72)[34]. Such lithium depletion is expected within a metal-poor first ascent red giant branch star[21] due to cycling of material in regions of high temperature where lithium is destroyed and subsequent dilution in an extensive convective envelope. This evolutionary status is in line with the adopted log *g* for SMSS 0313-6708.

**Derivation of upper limit on iron abundance.** An upper limit to the iron content has been calculated from a Markov Chain Monte Carlo analysis. The `emcee` package[35] was used, which is a Python[1] implementation of an affine-invariant ensemble sampler[36]. Two hundred walkers were used to explore the parameter space and maximise the log-likelihood. The likelihood function computes synthetic spectra using the MOOG code[37] and a 1-d model atmosphere[19] for SMSS 0313-6708 ($T_{eff}$ = 5,125 K, log $g$ = 2.3, [m/H] = -5, $v_t$ = 2 km/s) and compares it to portions of the normalised observed spectrum. The spectral regions surrounding the Fe lines at 385.9 nm and 371.9 nm were synthesised for different Fe abundances, smoothed with a Gaussian kernel (FWHM=9.6 km/s) to match the observations, and shifted to zero wavelength before being stacked. With logarithmic oscillator strengths of -0.431 and -0.710 respectively, these $\chi_{exc}$=0.0 eV transitions are the strongest Fe absorption lines available from our near-ultraviolet to infrared spectrum. The observed spectrum surrounding these Fe lines is similarly stacked with uncertainties added in quadrature based on the S/N of individual spectra, allowing for the $\chi^2$ difference between the model and observed spectra to be calculated. This model includes five free parameters: noise jitter, a constant scaling value for the stacked normalised spectrum, a residual velocity offset for each absorption line, and Fe/H. Using linear metallicity as a parameter (instead of logarithmic metallicity [Fe/H]) allows for both positive and approximate negative (i.e., emission spectra) column depths.

---

[1] http://www.python.org

A uniform prior between [0, 1] was employed for the noise jitter, and a uniformly distributed linear prior between [Fe/H] = -5.5 to -4.5 was adopted for metallicity. As the observed spectra is already at rest-frame and the laboratory wavelengths of these transitions are very well known, the velocity offset for each transition was treated as a Gaussian prior with $\mu$= 0 and sigma = 0.5 km/s. The continuum scaling factor prior was also treated as a Gaussian distribution, with $\mu$ = 1, $\sigma$ = 0.02. After 600 walker steps, an asymptotically-approaching mean acceptance fraction of 0.369 and a minimum reduced $\chi^2$ of 1.49 was observed. The resultant 67.8%, 95%, and 99.7% upper limits for [Fe/H]$_{1D,LTE}$ are <-7.7, <-7.5, <-7.3. This is demonstrated in Extended Data Figure 3, where stacked synthetic spectra for each metallicity is overlaid upon the stacked observed spectrum for the Fe lines employed. After corrections for 3-d and non-LTE effects, the 99.7% upper limit becomes [Fe/H]$_{3D,NLTE}$ = <-7.1.

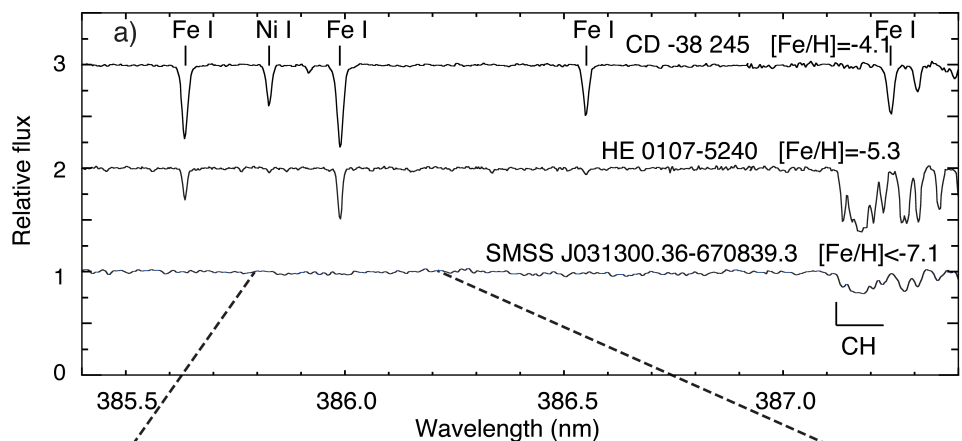
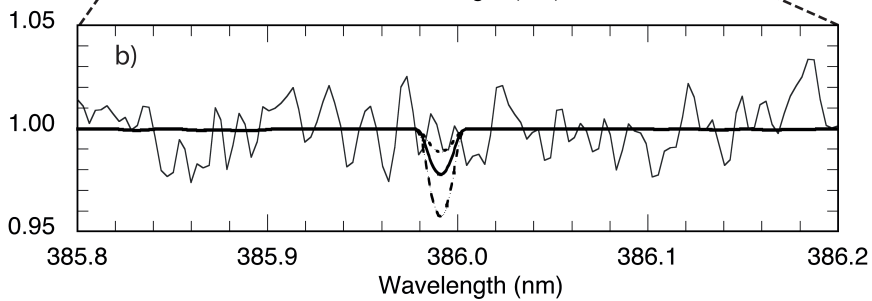
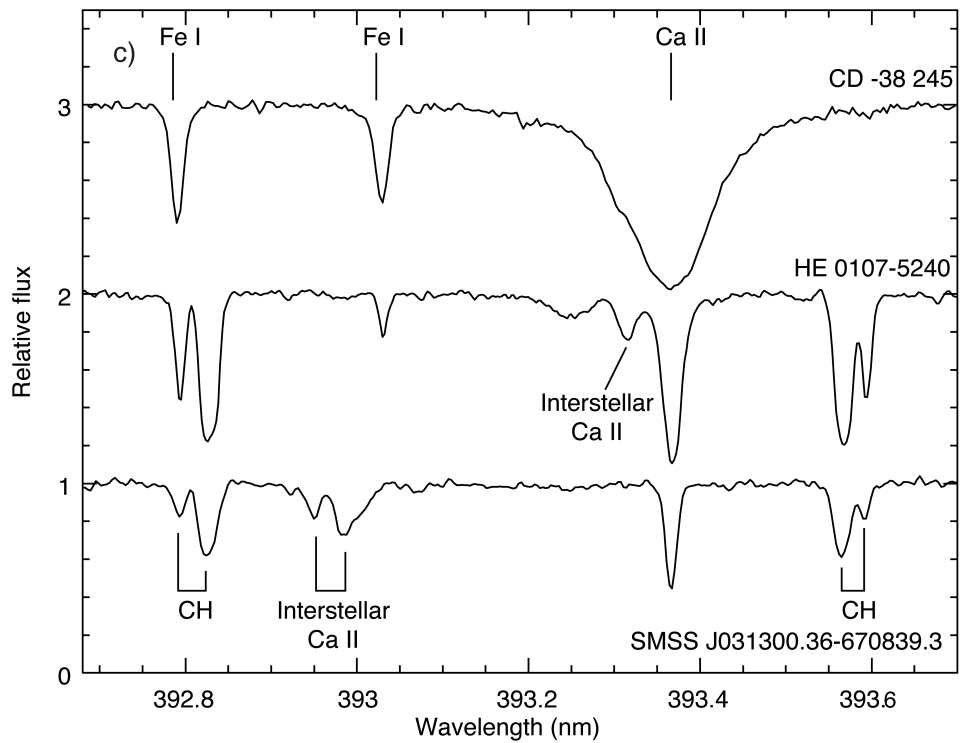

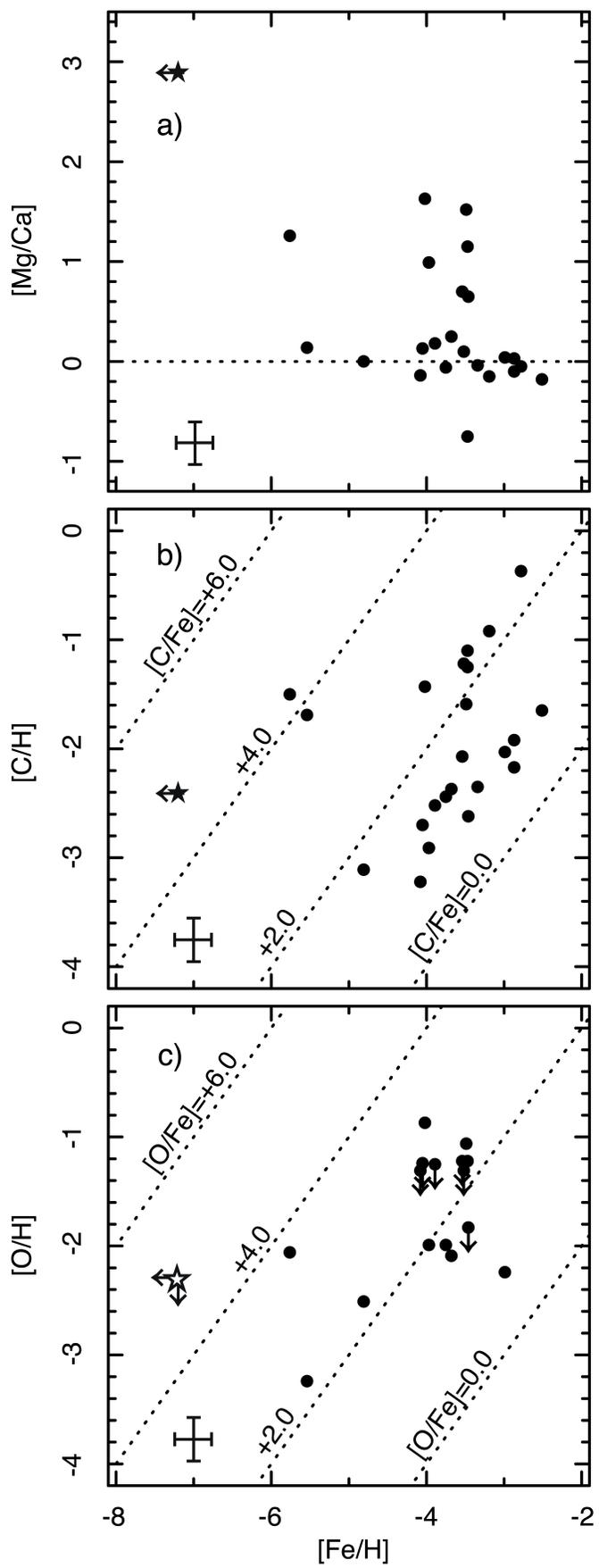

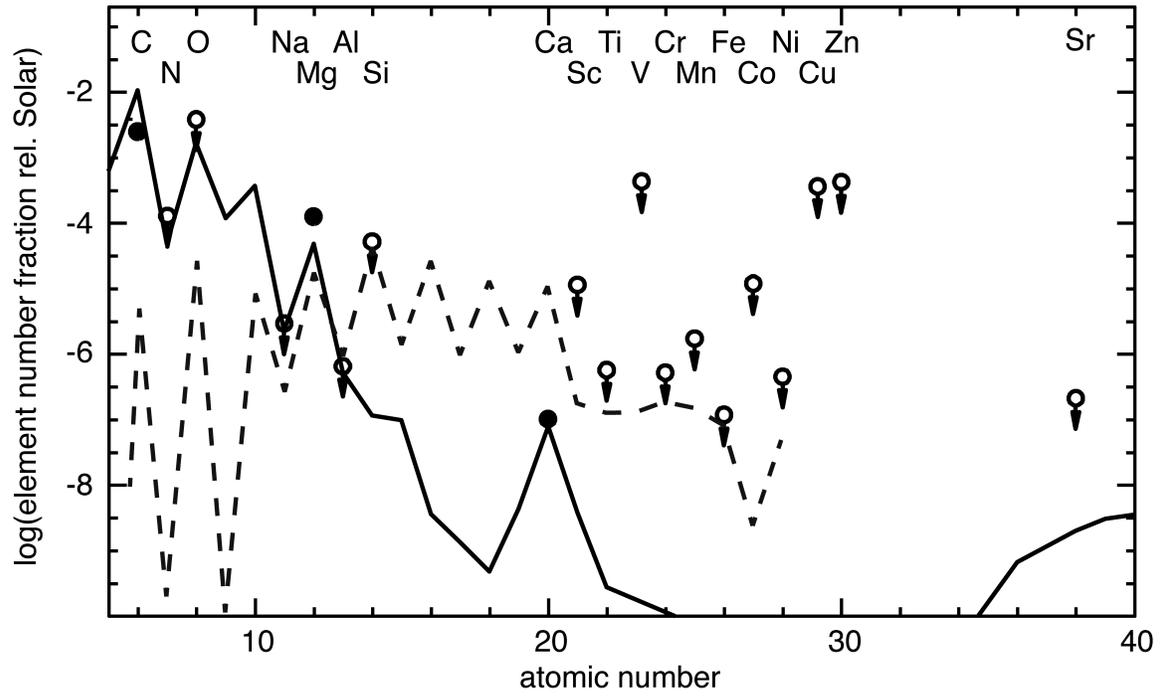

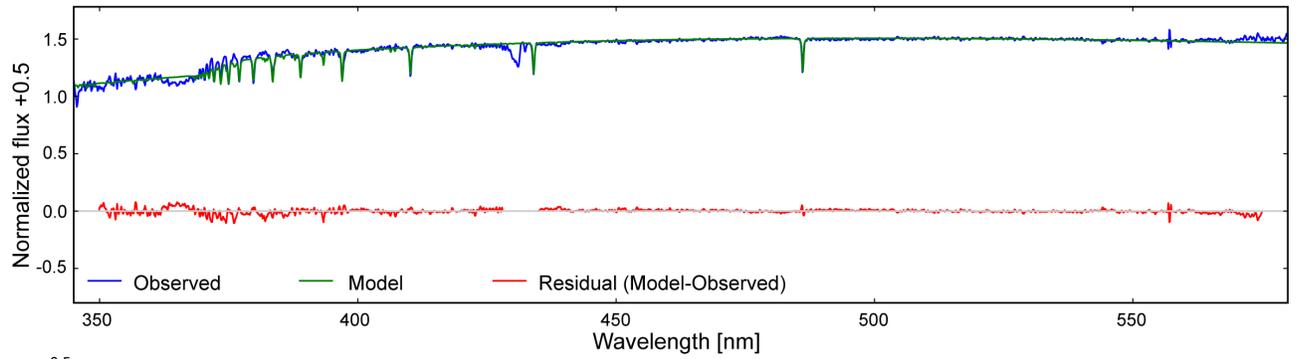
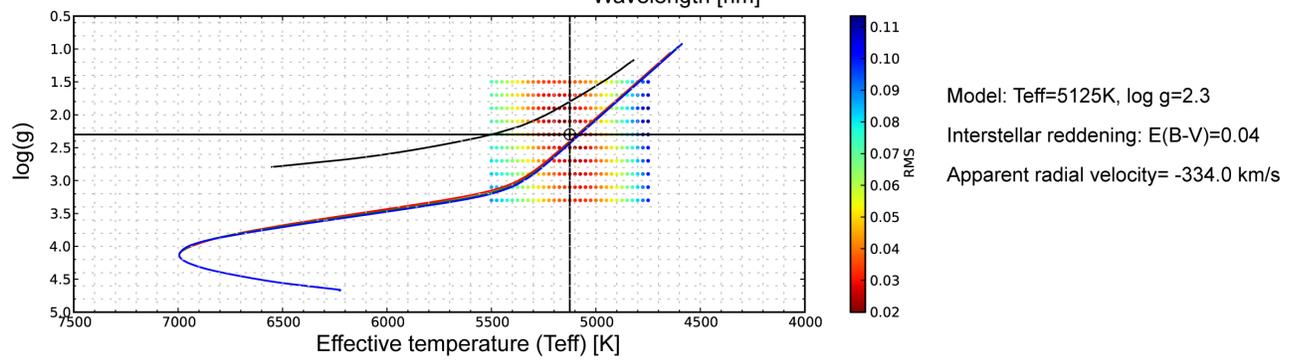

Model: Teff=5125K, log g=2.3

Interstellar reddening: E(B-V)=0.04

Apparent radial velocity= -334.0 km/s

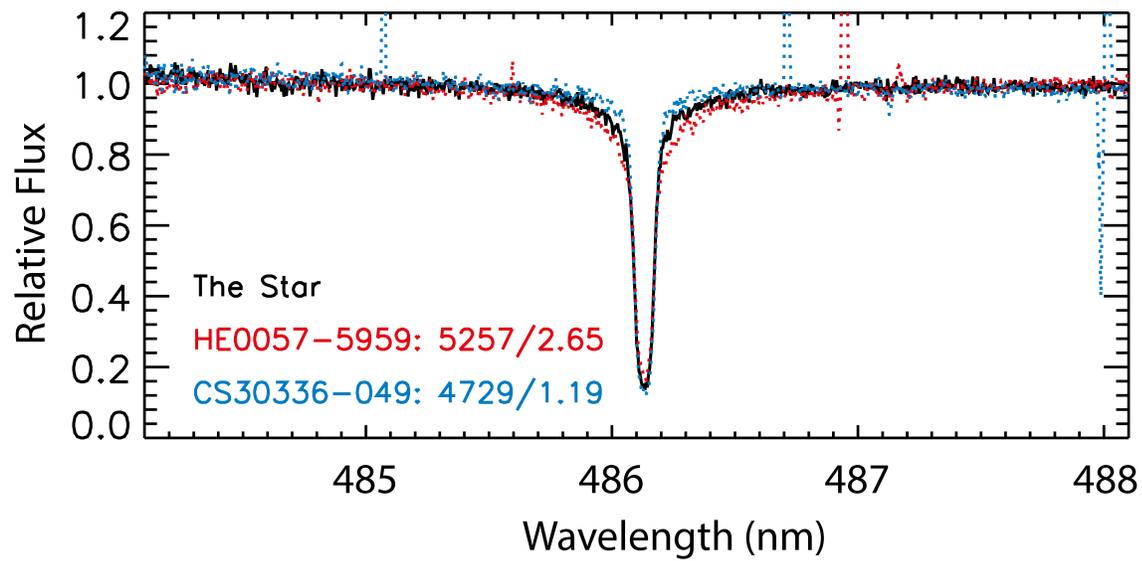
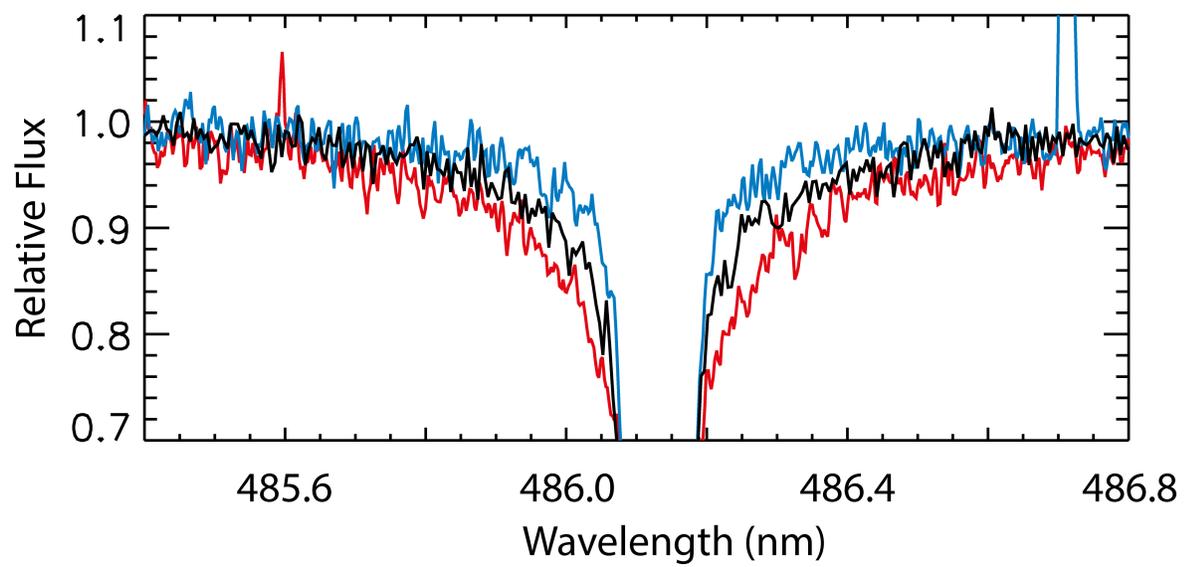

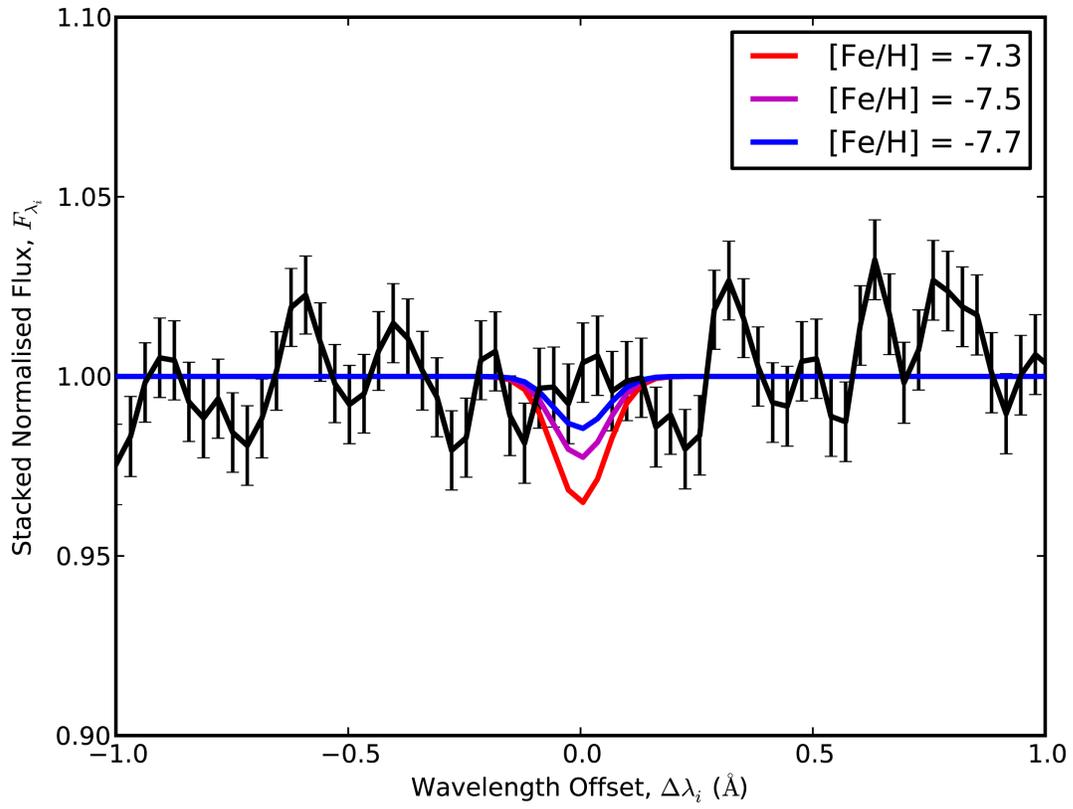